\documentclass[%
 reprint,
superscriptaddress,
 amsmath,amssymb,
 aps,
]{revtex4-2}

\usepackage{graphicx}
\usepackage{dcolumn}
\usepackage{bm}

\usepackage{float}
\usepackage{appendix}
\usepackage{graphicx}
\usepackage{hyperref}
\usepackage{physics}
\usepackage{dsfont}
\usepackage{xcolor}

\def\[#1\]{\begin{align}#1\end{align}}

\begin{document}

\preprint{APS/123-QED}

\title{Non-Gaussian entanglement revealed by higher-order quadrature cumulants}
\author{Abhinav Verma}
\email{abhve@dtu.dk}
\affiliation{Center for Macroscopic Quantum States (bigQ), Department of Physics, Technical University of Denmark, Fysikvej, 2800 Kongens Lyngby, Denmark}
\author{Olga Solodovnikova}\affiliation{Center for Macroscopic Quantum States (bigQ), Department of Physics, Technical University of Denmark, Fysikvej, 2800 Kongens Lyngby, Denmark}\affiliation{Center for Quantum Mathematics, University of Southern Denmark}
\author{Esben Klarlund}%
\affiliation{Center for Macroscopic Quantum States (bigQ), Department of Physics, Technical University of Denmark, Fysikvej, 2800 Kongens Lyngby, Denmark}
\author{Ulrik L. Andersen}%
\affiliation{Center for Macroscopic Quantum States (bigQ), Department of Physics, Technical University of Denmark, Fysikvej, 2800 Kongens Lyngby, Denmark}
\author{Jonas S. Neergaard-Nielsen}%
\email{jsne@fysik.dtu.dk}
\affiliation{Center for Macroscopic Quantum States (bigQ), Department of Physics, Technical University of Denmark, Fysikvej, 2800 Kongens Lyngby, Denmark}
\date{\today}

\begin{abstract}
Entanglement is central to quantum physics, yet detecting and exploiting it in non-Gaussian systems remains a major challenge. In continuous-variable platforms, standard inseparability criteria based on Gaussian statistics—such as the Duan–Simon criterion—fail when quantum correlations are encoded in higher moments of the field quadratures. Here we introduce a practical and scalable framework for detecting non-Gaussian entanglement using higher-order quadrature cumulants. In the Gaussian limit, the lowest-order condition reduces to the Duan–Simon criterion, while higher-order violations reveal entanglement inaccessible to second-order methods. We experimentally demonstrate tomography-free certification of non-Gaussian inseparability in photon-subtracted squeezed states of light, where Gaussian witnesses fail despite the presence of entanglement. 
This result identifies higher-order cumulants as natural and experimentally accessible observables for detecting non-Gaussian entanglement, and open new routes to harnessing non-Gaussian resources in continuous-variable quantum technologies.
\end{abstract}

\maketitle

\textbf{\textit{Introduction}}---
Entanglement is one of the defining features of quantum mechanics and underpins many emerging quantum technologies, including quantum communication, sensing, and computation. For continuous-variable systems, Gaussian states and operations provide a powerful and experimentally accessible framework: their correlations are completely described by second-order statistical moments of the quadrature operators. Within this setting, inseparability criteria such as those of Duan \textit{et al.} and Simon \cite{Duan,Simon} provide necessary and sufficient conditions for detecting Gaussian entanglement.

However, many quantum resources of practical and fundamental interest are intrinsically non-Gaussian, as such states are required for universal quantum computing and play a central role in long-distance quantum communication protocols. 
%
%
Despite substantial progress in understanding non-Gaussianity and the detection of entanglement in non-Gaussian states \cite{Duan,Simon,VanLoockFurusawa,entanglement_measure,Entanglement_detection,EPR,BellPaper,Metrological,nG_Quantification}, as well as in exploiting non-Gaussian resources for quantum technologies \cite{Tailor_nG_states, production_application_nG, gkp_breeding}, a scalable and experimentally practical method for probing non-Gaussian entanglement has been lacking. 
Full state tomography and approaches based on moment matrices \cite{PNAS2009, vogel2}  are experimentally demanding and computationally costly, limiting their scalability for studying entanglement in many physically relevant non-Gaussian states \cite{learningQuauntum, vogel1, Ys-Ra} especially as the state gets increasingly multimodal.

To formalize this limitation, we note that the Wigner function of any Gaussian state is fully determined by its first and second statistical moments. In this setting, the correlations contained in the covariance matrix  can always be brought to a canonical form via local symplectic transformations, allowing the separability properties of Gaussian states to be fully determined by covariance-based criteria. Consequently, for Gaussian states all entanglement information is contained at the level of second-order correlations. 

For non-Gaussian states, the situation is fundamentally different. There, correlations can reside in higher-order cumulants of the quadrature operators, which are not captured by the covariance matrix. A complete characterization of these states therefore requires an infinite hierarchy of cumulants. Importantly, these higher-order correlations cannot in general be removed or simplified through symplectic transformations, even when the covariance matrix itself appears separable. Thus, non-Gaussian entanglement may remain hidden from all criteria that rely solely on second-order moments (see Supplementary Information, section 1).


To overcome this limitation, we introduce and experimentally demonstrate a scalable, tomography-free framework for detecting inseparability based on higher-order quadrature cumulants \cite{fourth_moment_1,fourth_moment_2,vogel3}. Our approach extends the traditional covariance-based description by incorporating higher-order correlations, while reducing to the Duan–Simon condition in the Gaussian limit. The relevant cumulants can be directly extracted from quadrature statistics measured with standard homodyne and heterodyne detection, making the method experimentally accessible. Since the criterion relies on just a few homodyne and heterodyne measurements, it can be naturally extended to multimode systems in the same way as the van Loock-Furusawa criterion extends the Duan-Simon criterion to multiple parties \cite{VanLoockFurusawa}. This provides a significant practical advantage over full quantum state tomography, whose complexity grows exponentially with system size and quickly becomes prohibitive \cite{learningQuauntum}. 

\textbf{\textit{Inseparability of bipartite quantum states}}---
We start by providing an inseparability condition capable of detecting correlations that are not captured by covariance-based descriptions. Our approach extends the familiar continuous-variable entanglement criteria by incorporating higher-order cumulants of quadrature operators. To construct such a criterion, we consider a bipartite continuous-variable system and introduce two collective EPR-type quadrature operators
$\hat{u}=g_{1}\hat{x}_{1}+g_{2}\hat{x}_{2}$ and $\hat{v}=h_{1}\hat{p}_{1}+h_{2}\hat{p}_{2}$, where $g_i$ and $h_i$ are real coefficients that determine the relative contributions of the local quadrature operators $\hat{x}_i$ and $\hat{p}_i$. Such EPR-type combinations are widely used in continuous-variable entanglement criteria because they directly probe correlations between the two subsystems 

By analyzing correlations beyond second order, we show that separability imposes a constraint on the fourth-order cumulants of these collective observables. In particular, we prove (see Supplementary Information section 2 for the derivation) that any separable bipartite---generally mixed---state (though the inequality is stronger for pure states, see Supplementary information, section 2) must satisfy the following fourth-order truncated inequality:

\begin{widetext}
\begin{equation}
\begin{split}
    &W=\kappa_{4}(\hat{u})+\kappa_{4}(\hat{v}) +3 \kappa^{2}_{2}(\hat{u})+3 \kappa^{2}_{2}(\hat{v}) - 
    \sum_{i=1,2}\left[ \kappa_{2}^2(g_i\hat{x}_i) + \kappa_{2}^2(h_i\hat{p}_i) + T(\tilde{x}_i,\tilde{p}_i)\right]
    - 3|g_1h_1g_2h_2|\geq0
    \label{separability_equation_main}
\end{split}
\end{equation}
\end{widetext}
%
%
where 
$\tilde{x}_i=\hat{x}_i-\langle \hat{x}_i\rangle$, $\tilde{p}_i=\hat{p}_i-\langle \hat{p}_i\rangle$ are centralized quadrature operators and $\kappa_k(\hat{A})$ is the $k^\text{th}$ cumulant of operator $\hat{A}$. Cumulants provide a systematic way to characterize statistical correlations beyond Gaussian statistics: second-order cumulants correspond to variances and covariances, while higher-order cumulants quantify deviations from Gaussian behaviour. Intuitively, these quantities characterize the likelihood that quadrature variables undergo large excursions in phase space and the correlations between such excursions across the two subsystems. Gaussian states have vanishing cumulants above second order, whereas non-Gaussian states exhibit non-zero higher-order cumulants that encode correlations invisible to covariance-based descriptions.

The terms $T(\tilde{x}_i,\tilde{p}_i)$ contain intra-mode correlations between the two quadratures and are given by
\begin{equation} 
\begin{split}
    T(\tilde{x}_i,\tilde{p}_i)=2g_{i}^{2}h_{i}^{2}\sqrt{4|\text{Cov}(\tilde{x}_{i},\tilde{p}_{i})
    |^2+|\text{Cov}(\tilde{x}_{i}^{2} ,\tilde{p}_{i}^{2})|^{2}}
\end{split}
\end{equation}
where $\text{Cov}(\hat{A},\hat{B })= \frac{1}{2} \langle\{A,B\}\rangle-\langle A\rangle \langle B\rangle$ is the covariance of operators $\hat{A}$ and $\hat{B}$. In terms of cumulants, $T(\tilde{x}_i,\tilde{p}_i)$ directly depends on the joint cumulants $\kappa_{2,2}(\tilde{x}_i,\tilde{p}_i)$ and $\kappa_{1,1}(\tilde{x}_i,\tilde{p}_i)$. The values of these cumulants then determine the value of $T(\tilde{x}_i,\tilde{p}_i)$.

More generally, cumulants of arbitrarily high order can be included to construct a hierarchy of inseparability conditions that progressively capture increasingly complex non-Gaussian correlations. A brief discussion of this hierarchy is provided in Supplementary Information, section 2. Related completeness results have been shown in \cite{vogel2, PhysRevA.92.042328} using infinite-dimensional moment-matrix constructions and semidefinite optimization. However, such approaches rapidly become experimentally and computationally impractical as the number of modes or the order of moments increases. In contrast, the criterion introduced here involves only a finite number of experimentally accessible cumulants, making it well suited for scalable continuous-variable experiments.

\begin{figure}
\includegraphics[width=\columnwidth]{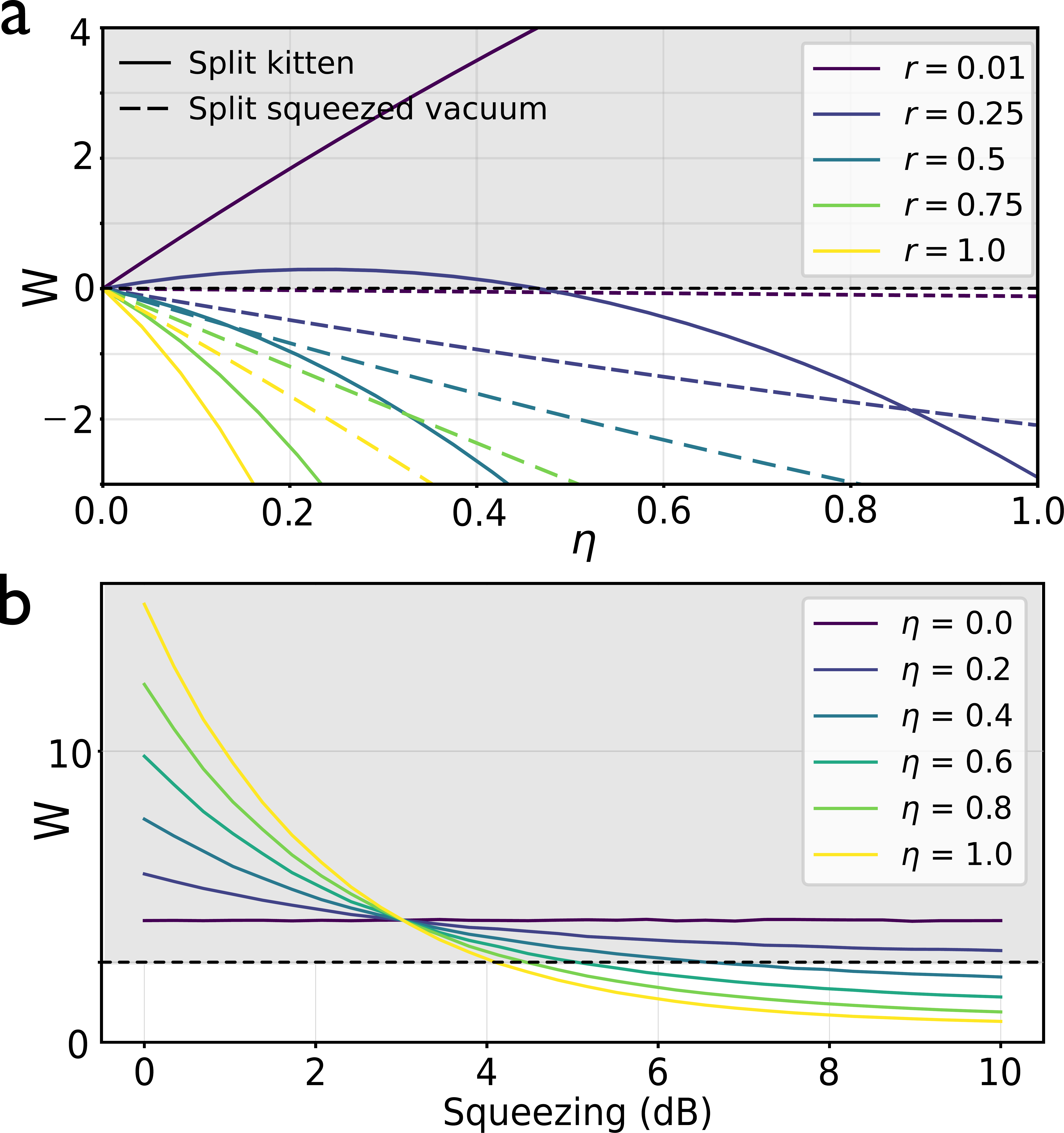}
    \caption{(\textbf{a}) Inseparability of a lossy split kitten as witnessed using the $4^{th}$ order criterion for different values of the squeezing parameter in contrast to its Gaussian counterpart - a lossy squeezed state split on a beam splitter with respect to losses. For very low squeezing, as the state approaches a single photon, the fourth order criterion loses traction and cannot detect inseparability require higher order conditions for certification. While entanglement persists, the criterion at this order fails and requires testing higher-order conditions. However, as the squeezing increases, the condition's faithfulness becomes increasingly robust to loss. The Gaussian state however remains strictly entangled as in the limit of Gaussians the criteria is at least as strong as that of Duan and Simon (see Supplementary information, section 6 for more details). (\textbf{b}) The inseparability criterion evaluated for the GKP Bell pair prepared by mixing two symmetrically lossy, finitely squeezed GKP qunaught states on a balanced beamsplitter vs. the squeezing of the peaks for different values of loss. The state is witnessed inseparable for some values of threshold squeezing which increases as the loss increases.}
    \label{fig:sims}
\end{figure}

that is subsequently split on a balanced beam splitter to generate bipartite entanglement. For comparison, we also show the Gaussian split squeezed vacuum state. The behaviour of the inseparability violation at different squeezing values are shown in Fig. \ref{fig:sims}a. At very low squeezing levels, the state approaches a single-photon state and is witnessed separable. Since the witness is sensitive only to fourth-order cumulants, the violation disappears even though entanglement actually persists \cite{PhysRevA.77.062333}. In this regime the remaining correlations reside in even higher-order cumulants requiring higher order tests. As loss removes the non-Gaussian features necessary for witnessing inseparability with the present criterion, the threshold also increases to counter the loss for detecting inseparability. In contrast, as squeezing level increases, the state is witnessed to be robustly inseparable across the full range of losses considered, with the transmissivity threshold smoothly decreasing toward zero. On the contrary, the Gaussian state is robustly seen inseparable, as expected, through the losses and all levels of squeezing as the criterion collapses to well known Gaussian criteria in the limit of Gaussian states (see supplementary information, section 6). Moreover, the criterion shows an increasingly higher level of violation for the non-Gaussian state as the squeezing level increases compared to the Gaussian state. This is also in agreement with previously observed increase in entanglement due to photon subtraction \cite{PS_entanglement,takahashi_entanglement_2010}.
 
 
 
 As a further test on practically relevant resource states, the criterion is evaluated in Fig. \ref{fig:sims}b for GKP Bell pairs, which are prime candidates for bosonic quantum error correction \cite{PRXQuantum.3.020334}. We simulate these as a mixture of two qunaught states on a beam splitter. Since loss rapidly degrades the non-Gaussian features in the GKP states, we observe - similar to the split photon-subtracted squeezed vacuum state - the appearance of squeezing thresholds that depend on the level of loss values. As losses increase, higher squeezing becomes necessary to witness inseparability at the fourth-order level. Interestingly, even in the lossless case, a finite squeezing threshold remains: below this value entanglement still exists, but no longer appears in the fourth-order correlations. Lastly, the extreme case of a completely lossy state remains separable for all values of squeezing for any arbitrary squeezing level. Another interesting artefact to notice is that the curves for different values of $\eta$ all intersect at $\epsilon=3dB$ where the value of $W=2$, which is the value the witness takes for the extreme, completely lossy state. This indicates the state start to become "more inseparable" than the fully lossy state at exactly this point irrespective of the loss. These behaviours collectively reflect the rich structure of the cumulant hierarchy associated with non-Gaussian states. 
 
 In both examples, the results indicate that detecting  inseparability in some cases require extending the witness to higher-order cumulants, thereby highlighting a fundamental limitation of any finite-order truncated hierarchy, particularly for strongly non-Gaussian states whose correlations may be distributed across many cumulant orders. Additional benchmark cases are discussed in the Supplementary Information.

\begin{figure}
    \centering
        \includegraphics[width=\columnwidth]{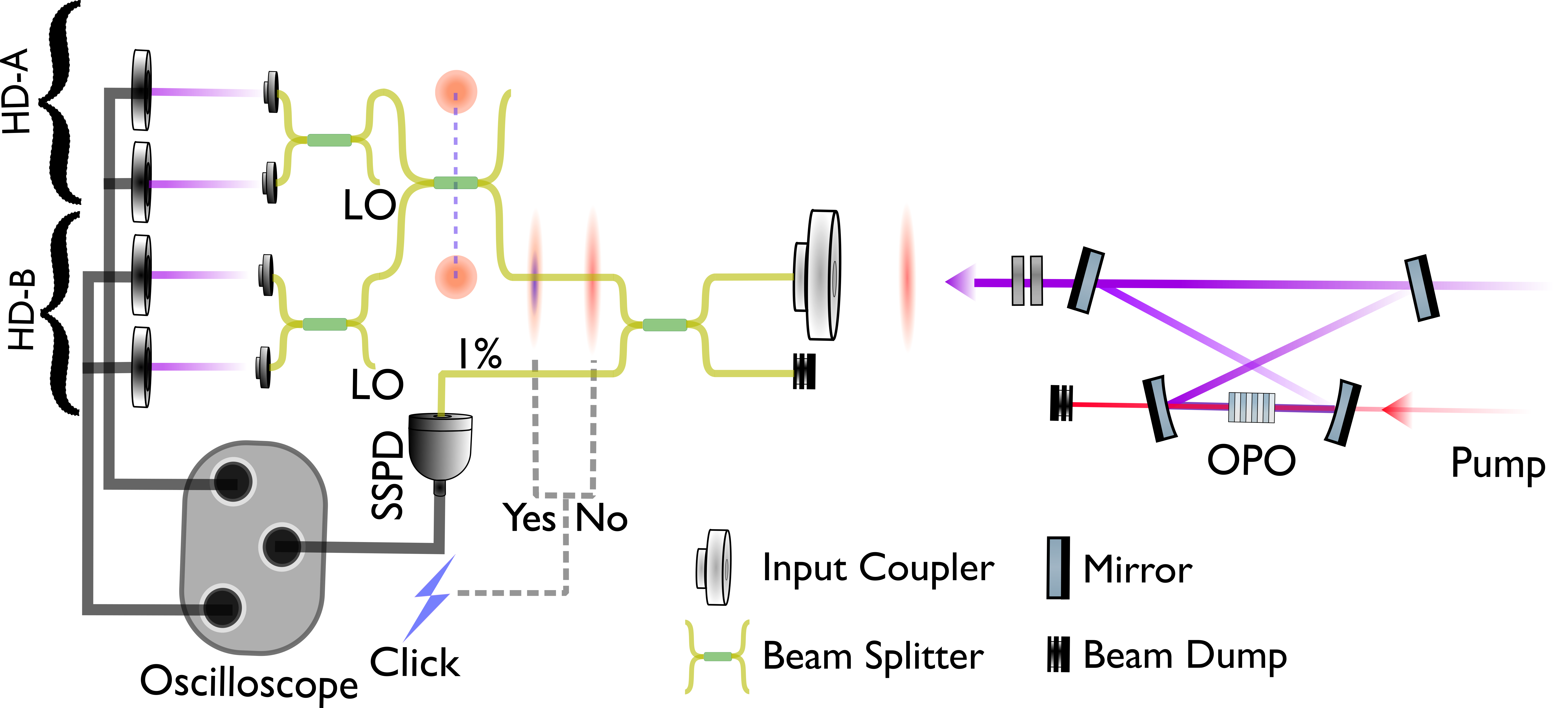}
    \caption{Experimental setup --- The kitten states are produced by performing photon subtraction on squeezed vacuum states. This is shown in the figure using a conditional preparation of detecting a 'click' on the superconducting single photon detector (SSPD). In the absence of the click, the state prepared is simply a squeezed vacuum. These states are then mixed with vacuum (or 'split') on a beam splitter before being detected with a pair of homodyne detectors over modes A and B while the modes' quadratures are measured on an oscilloscope for processing. The click serves as the trigger for identifying the photon-subtracted non-Gaussian mode. The statistics of the measurements are then used to verify the proposed inseparability conditions.}
    \label{fig:insep}
\end{figure}


\begin{figure*}
    \centering        
    \includegraphics[width=2\columnwidth]{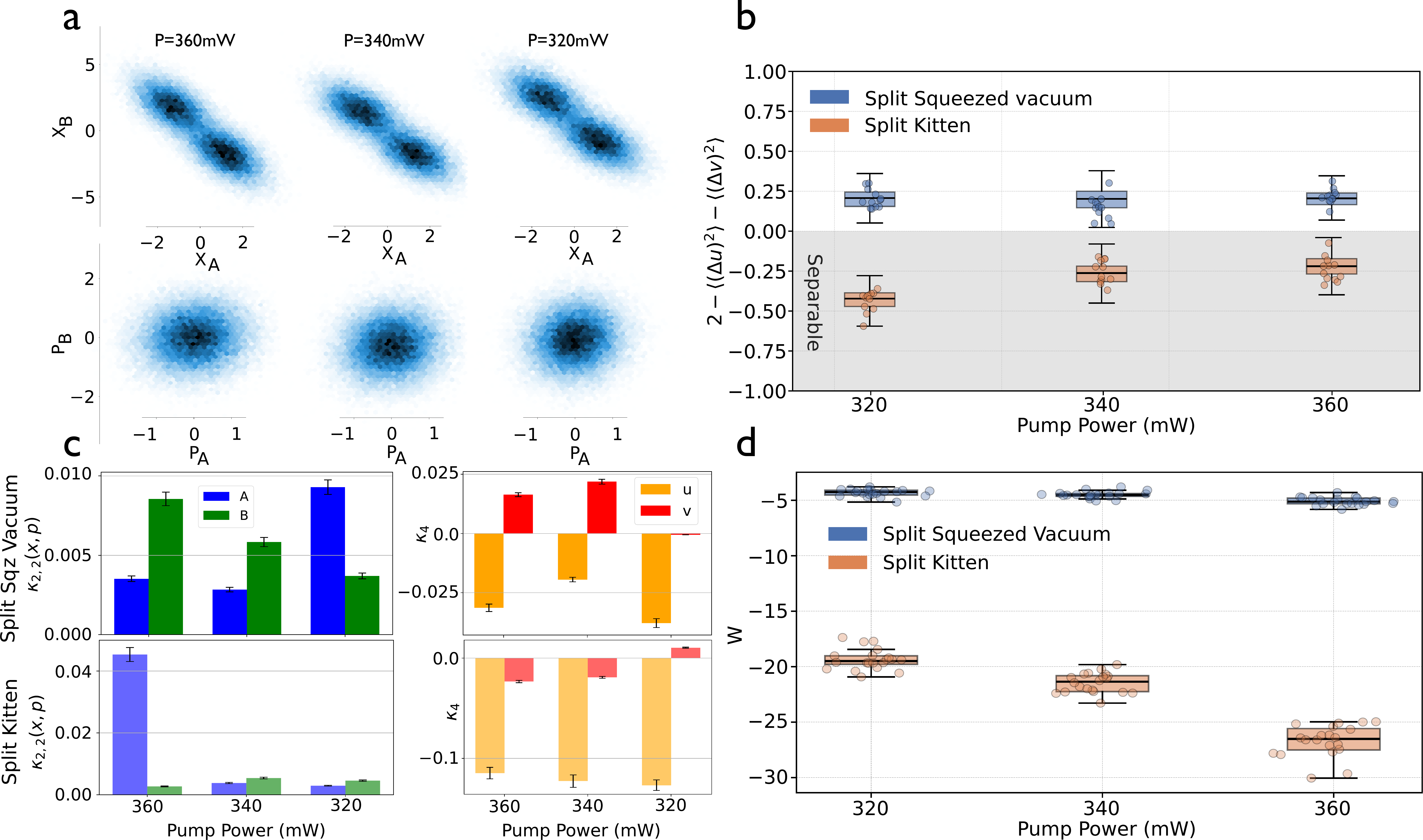}
    \caption{(\textbf{a}) Joint distributions of the kitten states obtained via homodyne measurements show clear non-Gaussianity in the x-marginals while the p-marginals show little non-Gaussianity. Both joint distributions, however, show increased correlations compared to their Gaussian counterparts - a signature of non-Gaussian entanglement (See Supplementary Information, section 7.6). The extent of violation by a squeezed vacuum state split on a beam splitter and a kitten state split on a beam splitter of  (\textbf{b}) the Duan criterion as well as (\textbf{d}) the inseparability criterion presented here computed from experimental data. The experiment was done for three pump powers as noted here --- 320 mW, 340 mW and 360 mW. Evidently there is violation of both criteria by the Gaussian state (split squeezed vacuum state) confirming inseparability.  Meanwhile the non-Gaussian state (the split kitten) only violates the non-Gaussian criteria presented in \eqref{separability_equation_main}. Panel (\textbf{c}) outlines the behaviour of pure and joint ($2,2$) $4^{th}$ order cumulants in the prepared states. These cumulants essentially vanish for Gaussian states while they indicate decreasing kurtosis values for the EPR operators with increasing values of pump power. This is well representative of the distributions show in panel (\textbf{a}).}
    \label{fig:vackitten}
\end{figure*}

\textbf{\textit{Experimental demonstration: Split kitten vs. split squeezed vacuum}} ---
To experimentally demonstrate the applicability and merit of the proposed inseparability criterion, we compare its performance on Gaussian and non-Gaussian states generated in the laboratory. Specifically, we investigate the two types of bipartite states studied in Fig.~\ref{fig:sims}, namely those obtained by splitting on a balanced beamsplitter a Gaussian squeezed vacuum state or a non-Gaussian photon-subtracted squeezed state (kitten). The experimental preparation of these states is illustrated schematically in Fig. \ref{fig:insep}. 
Squeezing at 1550 nm wavelength is generated in an optical parametric oscillator (OPO). After coupling to fiber, 1\% of the light is directed to a superconducting nanowire single-photon detector (SSPD), while the remaining light is split 50:50 and measured on homodyne detectors. The homodyne signals are subsequently filtered by a temporal mode centered around the arrival time of SSPD ``clicks''. Further details of the state preparation and data analysis procedures are provided in the Supplementary Information, Sec.~7. 

The resulting joint quadrature distributions are obtained by simultaneously measuring $x$ and $p$ quadratures of the two output modes using homodyne detection. To obtain the values of $T(\tilde{x_i},\tilde{p_i})$ for the $i^{th}$ mode we use heterodyne detection. Since the vacuum mode that mixes in the ensuing heterodyne detection is Gaussian, all cumulant contribution of order greater than $2$ vanishes (see Supplementary Information, section 3). Thus the presence of non-zero $\kappa_{2,2}$, $\kappa_{4}$ can only come from the non-Gaussian nature of the quantum states.  The experimentally obtained joint distributions are shown in Fig. \ref{fig:vackitten}a for three slightly different powers of the OPO pump. These quadrature measurements allow the cumulants appearing in the inseparability criterion (\ref{separability_equation_main}) to be extracted directly. 
%
The joint $x_A, x_B$ quadrature distributions are clearly non-Gaussian.
In contrast, the corresponding $p_A,p_B$ distributions remain approximately Gaussian, reflecting the phase-sensitive nature of the photon-subtraction process.

These observations are quantitatively supported by the extracted cumulants. The fourth-order cumulants $\kappa_{4}(\hat{u})$ and $\kappa_{4}(\hat{v})$ (where we use $g_1=-g_2=1$ and $h_1=h_2=1$), shown in Fig.~\ref{fig:vackitten}c, clearly deviate from zero for the non-Gaussian states while remaining close to zero for the Gaussian squeezed vacuum. The non-Gaussian states exhibit pronounced negative kurtosis, indicating sub-Gaussian statistics with heavier tails than a Gaussian distribution. The magnitudes decrease with increasing pump power, reflecting the diminishing strength of higher-order correlations as the states get more mixed and lose Wigner negativity \cite{neergaard-nielsen_generation_2008}. 
The joint cumulants $\kappa_{2,2}(\hat{x}_A,\hat{x}_B)$ and $\kappa_{2,2}(\hat{p}_A,\hat{p}_B)$ remain small, whose distributions can be seen in the Supplementary Information.


Figures~\ref{fig:vackitten}b and \ref{fig:vackitten}d show the resulting evaluation of the inseparability criterion. As the non-Gaussian character of the state increases, the violation of the cumulant-based criterion becomes significantly stronger compared to the Gaussian reference states. 
Importantly, Fig.~\ref{fig:vackitten}b highlights a regime where the standard Duan inseparability witness fails to detect entanglement in the photon-subtracted squeezed state, while the cumulant-based witness successfully certifies inseparability. This occurs because the higher-order cumulants capture correlations that lie beyond the covariance matrix description assumed in second-order criteria.

For comparison, the Gaussian reference state—obtained by splitting a squeezed vacuum on a balanced beam splitter—also exhibits entanglement. However, all cumulants above second order vanish for this Gaussian state, and the inseparability can therefore be detected using second-order criteria. In contrast, the non-Gaussian state exhibits substantial fourth-order pure cumulants, indicating that a significant fraction of the correlations resides in higher-order statistics. At the same time, the heavier tails and sub-Gaussian statistics of the photon-subtracted state lead to increased quadrature variances, which can mask entanglement when evaluated using second-order witnesses. This explains why the entanglement of the split-kitten state becomes invisible to traditional covariance-based criteria, as illustrated in Fig.~\ref{fig:vackitten}b.

Taken together, these results demonstrate that the cumulant-based inseparability criterion provides a practical and experimentally accessible method for detecting both Gaussian and non-Gaussian entanglement. In particular, it reveals correlations residing in higher-order statistics that remain completely hidden to conventional second-order tests, thereby providing a powerful diagnostic tool for non-Gaussian continuous-variable quantum states.


\textit{Conclusion}
We have introduced and experimentally demonstrated a practical and scalable method for detecting non-Gaussian entanglement in continuous-variable systems. By exploiting higher-order quadrature cumulants, the proposed framework reveals correlations that remain invisible to covariance-matrix-based criteria. In contrast to many existing approaches, the method requires neither full quantum state tomography nor infinite moment hierarchies, and relies only on experimentally accessible quadrature measurements. Beyond its immediate utility as a detection tool, the cumulant-based framework provides a new perspective on the structure of quantum correlations in non-Gaussian states and their applications to improving performance in well known protocols. It enables systematic exploration of non-Gaussian resources that underpin, for example, long-distance quantum communication and quantum error correction.

Looking ahead, extending this framework to large-scale multimode photonic systems—such as cluster states used for measurement-based quantum computing—represents an important future direction. The approach may therefore provide a practical tool for characterizing non-Gaussian entanglement in next-generation photonic quantum processors.

\textbf{\textit{Aknowledgements}}---We acknowledge support from the Danish National Research Foundation (bigQ, DNRF0142), Innovation Fund Denmark (QuantERA - ClusSTAR, 3155-00024A), EU ERC Adv (ClusterQ  grant no. 101055224) and EU Horizon Europe (CLUSTEC, grant agreement no. 101080173).

\bibliography{apssamp}
\begin{appendices}
    
\end{appendices}

\end{document}